\documentclass[12pt,letterpaper]{article}

\usepackage{natbib}
\usepackage{hyperref}
\usepackage[left=1in, top=1in, right=1in, bottom=1in]{geometry}
\usepackage{graphicx}
\usepackage{bm}
\usepackage{colonequals}
\usepackage{amsmath}
\usepackage{amssymb}
\usepackage{dsfont}
\usepackage{url}
\usepackage[dvipsnames]{xcolor}
\usepackage{bbm}
\usepackage{array}
\usepackage{tabularx}
\usepackage{multirow}
\usepackage[font={footnotesize}]{caption,subcaption}
\usepackage[utf8]{inputenc}
\usepackage{enumitem}
\usepackage{soul} %
\usepackage{placeins} %
\usepackage[normalem]{ulem}
\usepackage{algorithm2e}

\graphicspath{{plots/}}

\usepackage{xr}
\makeatletter
\newcommand*{\addFileDependency}[1]{%
  \typeout{(#1)}
  \@addtofilelist{#1}
  \IfFileExists{#1}{}{\typeout{No file #1.}}
}
\makeatother
\newcommand*{\myexternaldocument}[1]{%
    \externaldocument{#1}%
    \addFileDependency{#1.tex}%
    \addFileDependency{#1.aux}%
}
\myexternaldocument{supplement}

\usepackage{mathtools}
\mathtoolsset{showonlyrefs}

\bibpunct[, ]{(}{)}{;}{a}{,}{,}

\newcommand{\bb}{\bm{b}}
\newcommand{\bd}{\bm{d}}
\renewcommand{\bf}{\bm{f}}

\newcommand{\bs}{\bm{s}}

\newcommand{\by}{\bm{y}}

\newcommand{\bS}{\bm{S}}

\newcommand{\bY}{\bm{Y}}

\newcommand{\bE}{\bm{E}}

\newcommand{\bD}{\bm{D}}

\newcommand{\bU}{\bm{U}}
\newcommand{\bV}{\bm{V}}
\newcommand{\bK}{\bm{K}}

\newcommand{\bQ}{\bm{Q}}

\newcommand{\bR}{\bm{R}}

\newcommand{\bftheta}{\bm{\theta}}

\newcommand{\bfkappa}{\bm{\kappa}}

\newcommand{\bfpsi}{\bm{\psi}}

\newcommand{\bfzeta}{\bm{\zeta}}

\newcommand{\bfLambda}{\bm{\Lambda}}

\newcommand{\GP}{\mathcal{GP}}

\newcommand{\normal}{\mathcal{N}}

\DeclareMathOperator*{\argmin}{arg\,min}
\DeclareMathOperator*{\argmax}{arg\,max}

\title{Bayesian nonparametric generative modeling of large multivariate non-Gaussian spatial fields}

\author{Paul F.V.\ Wiemann\thanks{Department of Statistics, Texas A\&M University\\\phantom{asdasf}3143 TAMU, College Station, TX 77843} \and Matthias Katzfuss\footnotemark[1] \thanks{Corresponding author: \texttt{katzfuss@gmail.com}}}
\date{}

\begin{document}
\maketitle

\begin{abstract}
\noindent
Multivariate spatial fields are of interest in many applications, including climate model emulation. Not only can the marginal spatial fields be subject to nonstationarity, but the dependence structure among the marginal fields and between the fields might also differ substantially.
Extending a recently proposed Bayesian approach to describe the distribution of a nonstationary univariate spatial field using a triangular transport map, we cast the inference problem for a multivariate spatial field for a small number of replicates into a series of independent Gaussian process (GP) regression tasks with Gaussian errors.
Due to the potential nonlinearity in the conditional means, the joint distribution modeled can be non-Gaussian.
The resulting nonparametric Bayesian methodology scales well to high-dimensional spatial fields.
It is especially useful when only a few training samples are available, because it employs regularization priors and quantifies uncertainty. 
Inference is conducted in an empirical Bayes setting by a highly scalable stochastic gradient approach.
The implementation benefits from mini-batching and could be accelerated with parallel computing. 
We illustrate the extended transport-map model by studying hydrological variables from non-Gaussian climate-model output.

\end{abstract}


\section{Introduction \label{sec:intro}}

Multivariate spatial fields play a significant role in various scientific disciplines, including environmental modeling and climate science, where multiple spatially referenced variables are observed.
To highlight the need for statistical models to effectively capture the intricate relationships among multiple variables in spatial fields, consider measures for apparent temperature.
The heat index, for instance, relies on temperature and humidity to assess the apparent temperature. More comprehensive measures of apparent temperature, such as the wet-bulb globe temperature, which incorporates the additional factors of wind speed and radiation, have been proposed to better assess the stress of exposure to high temperatures on the human body. Consequently, in the context of statistical climate-model emulators, capturing the inter-variable dependence is crucial for reliable predictions, conditional predictions, and accurate uncertainty quantification. Accurately inferring the joint distribution and understanding the conditional relationships among these variables is challenging, especially when dealing with complex dependencies and non-Gaussian characteristics.

Most existing methods for univariate or multivariate spatial analysis were developed for inference based on a single training sample and assume Gaussian processes (GPs) with simple parametric covariance functions \citep[e.g.,][]{Cressie1993, Banerjee2004}. Extensions to non-parametric covariances \citep[e.g.,][]{Huang2011, Choi2013, Porcu2019} or multivariate fields \citep[e.g.,][]{Genton2015} typically still rely on implicit or explicit assumptions of Gaussianity.
For emulation of univariate spatial climate-model output, one can combine locally fit anisotropic Mat\'ern covariances into a global Gaussian model \citep{Nychka2018, Wiens2020}.
Generative machine-learning approaches \citep[e.g.,][]{kobyzev2020normalizing, Goodfellow2016, Kovachki2020} often require many training samples and may be sensitive to tuning-parameter and network-architecture choices \citep[e.g.,][]{Arjovsky2017, Hestness2017, Mescheder2018}.

Spatial-temporal dependencies of multivariate global fields were captured by \citet{Jun2011}. Nonstationarities in latitude are supported by \citet{Castruccio2013} for the univariate field describing annually averaged surface temperature. \citet{Edwards2019} extend this method to multivariate fields relying on a parametric approach and assuming a Gaussian distribution.
Relying on a stationary assumption while allowing for arbitrary missingness patterns, \citet{guinness2022nonparametric} estimates the spectral form of gridded multivariate spatio-temporal data.
In contrast to the method we propose, these multivariate models required all variables to be observed at the same spatial locations.

Triangular transport maps \citep[e.g.,][]{Marzouk2016} can be used to characterize continuous multivariate distributions. A transport map transforms the target distribution into a reference distribution, such as the standard Gaussian. Non-Gaussian target distributions can be obtained by introducing nonlinearities into the map. With an invertible transport map, one can sample from the target distribution and its conditionals or convert non-Gaussian data to the reference space, where linear regression or interpolation can be applied.
Transport maps are often estimated from training data by iteratively expanding a finite-dimensional parameterization of the transport map \citep[e.g.,][]{ElMoselhy2012, Bigoni2016, Marzouk2016, Parno2016}. 

\citet{Katzfuss2021} instead proposed a Bayesian nonparametric approach, in which the components of the transport map are modeled as GPs. This results in closed-form inference that quantifies uncertainty and avoids under- and over-fitting even when the number of training samples is small. For target distributions corresponding to spatial fields, \citet{Katzfuss2021} proposed specific priors that exploit the screening effect via suitable conditional-independence assumptions that guarantee computational scalability for very large datasets. The resulting sparse non-linear transport maps can be seen as a non-parametric and non-Gaussian generalization of Vecchia approximations \citep[e.g.,][]{Vecchia1988, Stein2004, Datta2016, Katzfuss2017a, Schafer2020}, which implicitly utilize linear transport map given by a sparse inverse Cholesky factor. \citet{Kidd2020} proposed a Bayesian non-parametric inference on the Cholesky factor.

Our contribution is a novel extension of the scalable Bayesian transport map (BTM) approach developed by \citet{Katzfuss2021}, tailored explicitly for learning the distribution of multivariate spatial fields from a few replicates.
The essential contribution lies in the introduction of an augmented input space that incorporates both the spatial locations and latent locations referencing the component from the multivariate response \citep[see][for a similar concept used with a stationary parametric covariance function]{Apanasovich2010LatentDimensions}.
By leveraging the augmented input space, we expand the scope of BTM to encompass multivariate spatial fields without fundamentally altering the core principles and estimation algorithms.
Consequently, the extension benefits from the good approximation properties, making the method scale well to very large spatial data sets while being trainable from a small number of replicates.

The remainder of the paper is organized as follows. 
In Section~\ref{sec:methodology}, we provide an overview of the methodology, including a review of Bayesian transport maps, the proposed extension estimation procedures, and computational considerations.
Section~\ref{sec:sim-study} presents numerical comparisons in a simulation study.
In Section~\ref{sec:application}, we discuss the results of an application to climate-model output. 
Finally, we conclude in Section~\ref{sec:conclusions}, summarizing the contributions, discussing future directions, and emphasizing the importance of our findings.

\section{Methodology \label{sec:methodology}}

Consider a $P$-variate spatial field, where the locations at which the observations are available might differ among the marginal fields. 
Let $y(p, \bs)$ denote the random variable associated with the $p$-th marginal field at location $\bs \in \mathcal{S}$, where $\mathcal{S}$ is the spatial domain and $p=1,\dots, P$.
Moreover, let $\by_p = (y(p, \bs^{(p)}_1), \dots, y(p, \bs^{(p)}_{N_p}))^T$ be the random vector associated with the $p$-th spatial field observed at locations $\bs^{(p)}_1, \dots, \bs^{(p)}_{N_p} \in \mathcal{S}$.
Finally, $\by = (\by_1^T, \dots, \by_P^T)^T$ combines the marginal fields by concatenating $\by_1, \dots, \by_P$.

We would like to infer the distribution of $\by$ from $R$ independent replicates denoted as $\mathbf{Y} = \by^{(1)}, \dots, \by^{(R)}$, drawn from the same distribution as $\by$.
The scalable Bayesian transport map method, proposed by \citet{Katzfuss2021}, provides an approach to infer the distribution of a univariate field.
The method focuses on modeling the dependence structure using a nonparametric approach assuming $\by$ to be a field with zero expectation.
In this section, we review the existing methodology and propose an extension to multivariate fields.

\subsection{Review of the scalable Bayesian transport map}

Throughout this subsection, we assume $P = 1$ and drop the corresponding index for simplicity.
A transport map $\mathcal{T}: \mathds{R}^N \to \mathds{R}^N$ is a map that characterizes the distribution of $\by$ by providing a transformation of $\by$ to a simple reference distribution, e.g., $\mathcal{T}(\by) \sim \normal(\bm{0}, \bm{I}_N)$. Without loss of generality, the map can have a lower triangular form \citep{Rosenblatt1952, Carlier2009} such that
\begin{align*}
	\mathcal{T}(\by) = \begin{pmatrix}
		\mathcal{T}_1(y_1)\\
		\mathcal{T}_2(y_1, y_2)\\
		\vdots\\
		\mathcal{T}_N(y_1, \dots, y_N)
	\end{pmatrix}
\end{align*}
with $\mathcal{T}_n$ being strictly monotone in the $n$-th argument.
\citet{Katzfuss2021} model $\mathcal{T}_n(\by_{1:n}) = (y_n - f_n(\by_{1:n-1}))d_n^{-1}$, 
where $\by_{i:j} = (y_i, \dots, y_j)^T$ and $\bd = (d_1, \dots, d_N)^T$, and $\bf = (f_1, \dots, f_N)^T$ are random vectors.
This facilitates the factorization of the joint distribution of $\by$ as
\begin{equation}
p(\by)
= p(y_1) \prod_{n=2}^{N} p(y_n | y_1, \dots, y_{n-1})
= \prod_{n=1}^N \int  \normal(y_n | f_n(\by_{1:n-1}),d_n^2)p(f_n, d_n) \, \text{d}d_n\, \text{d}f_n,
\label{eq:lik-fact}
\end{equation}
to which we refer as the integrated likelihood.

\subsubsection{Priors}
An independent Gaussian-process inverse-Gamma prior is placed on each pair $(f_n,d_n^2)$ for $n=1, \dots, N$:
\begin{align*}
    d_n^2 &\sim \mathcal{IG}(\alpha_n,\beta_n), &&\text{with } \alpha_n>1, \; \beta_n > 0\\
    f_n | d_n &\sim \GP(0,d_n^2 K_n) &&\text{with covariance function } K_n.
\end{align*}
The priors' parameters, including parameters determining $K_n$ here referred to as $\bfkappa_n$, $\bfpsi = \{\alpha_1, \dots, \alpha_N, \beta_1, \dots, \beta_N, \bfkappa_1, \dots, \bfkappa_N\}$ may depend on a hyperparameter vector $\bftheta$. In the spatial case described by \citet{Katzfuss2021} as well as in our multivariate extension, $\bfpsi$ deterministically depends on $\bftheta$. Moreover, the dimensionality of $\bftheta$ is much smaller than that of $\bfpsi$, as the number of hyperparameters does not increase with $N$.

Given the substantial number of random variables involved, it becomes imperative to select the prior parameters judiciously. These priors are formulated based on the principles of shrinkage and theoretical considerations pertinent to Gaussian processes with specific covariance functions.

To provide more clarity, we introduce some additional notations. We denote $o(n)$ as a sequence that orders the values preceding $y_n$ by increasing the distance of their associated locations to $\bs_n$. Formally, $o(n)$ is the sequence such that for $1 \leq i < j < n$, it holds that $\Vert \bs_n - \bs_{o(n)_i}\Vert \leq \Vert \bs_n - \bs_{o(n)_j}\Vert$. Additionally, we define $\ell_n$ as the minimum distance from $\bs_n$ to any of its preceding neighbors, specifically as $\ell_n = \min_{i \in {1, \dots, n-1}} \Vert \bs_n - \bs_{i}\Vert = \Vert \bs_n - \bs_{o(n)_1}\Vert$.

Based on the reasoning of \citet{Schafer2017}, \citet{Katzfuss2021} observe a roughly power-functional decay of the conditional variance $\mathrm{Var}[y_n | f_n, \by_{1:n}] = d_n^2$ with respect to the distance to the nearest neighbor $\ell_n$.
The prior on $d_n^2$ captures this relationship by incorporating hyperparameter values $\alpha_n$ and $\beta_n$ derived from the condition that the prior expectation and prior standard deviation of $d^2_n$  should satisfy $\mathrm{E}[d_n^2] = \exp(\theta^d_1 + \exp(\theta^d_2) \log(\ell_n))$ and $\mathrm{SD}[d_n^2] = g\mathrm{E}[d_n^2]$ with constant $g > 0$, respectively. A relatively weak prior is constructed by setting $g=4$. The number of hyperparameters in the priors of $d_n^2$ that need to be estimated reduces from $2N$ to $2$.

The inference of $f_n$ becomes feasible through the application of two key principles. Firstly, as the distance to the nearest neighbor, denoted as $\ell_n$, decreases, the prior exerts a stronger tendency to push $f_n$ towards linearity. Secondly, drawing motivation from the ``screening effect'' \citep{Stein2011} and the work by \citet{Schafer2017}, the inputs to $f_n$ are regularized such that the relevance of each input diminishes exponentially according to its position in the ordered input sequence, namely, $\by_{o(n)} = (y_{o(n)_1}, \dots, y_{o(n)_{n-1}})^T$.
These principles lead to the covariance function
\begin{align*}
	K_n(\by_{o(n)}, \by_{o(n)}') %
	&=(\mathrm{E}[d^2_n])^{-1}\left(\by_{o(n)}^T\bQ_n \by_{o(n)}' + \sigma^2_n\rho \left(\frac{\sqrt{(\by_{o(n)} -  \by_{o(n)}')^T \bQ_n (\by_{o(n)} -  \by_{o(n)}')}}{\gamma}\right)\right).
\end{align*}
Here, $\bQ_n = \operatorname{diag}_j(\exp(-j\exp(\theta^q)))$ encodes the decreasing relevance of more distant inputs, $\rho$ represents the Matérn correlation function with three-halves smoothness, and $\gamma = \exp(\theta^{\gamma})$ serves as a range parameter. The parameter $\sigma_n$ governs the nonlinearity of $f_n$ and, a priori, decays in a manner similar to $\mathrm{E}(d^2_n)$, specifically as $\sigma^2_n = \exp(\theta^\sigma_1 + \exp(\theta^\sigma_2) \log(\ell_n))$.
The hyperparameters that determine the values of $\bfpsi$ are encompassed within the hyperparameter vector $\bftheta = (\theta^q, \theta^\gamma, \theta^d_1, \theta^d_2, \theta^\sigma_1, \theta^\sigma_2)^T$.

\subsubsection{Vecchia approximation and hyperparameter estimation}
To enable inference for small training sizes $R$ and to ensure scalability to large spatial fields $N$, $f_n$ is assumed to depend only on a conditioning set of restricted size.
Concretely, $f_n(\by_{1:n-1})$ is replaced by $f_n(\by_{c(n)})$, where $c(n) \subseteq \{1, \dots, n-1\}$ with $|c(n)| = \min(m, n-1)$ and $\by_{c(n)}$ is the subvector of $\by_{1:n-1}$ with the indices found in $c(n)$.
The good approximation properties of the reduced conditioning set are achieved by ordering the vector $\by$ according to the maximum-minimum (maxmin) ordering and then selecting the nearest neighbors for the reduced conditioning set~$c(n)$.
For a comprehensive discussion of different orderings, refer to \citet{Guinness2016a}.
In the case of the scalable Bayesian transport map, the computational complexity to determine the $n$-th element of the posterior map reduces from $\mathcal{O}(R^3 + nR^2)$ to $\mathcal{O}(R^3 + mR^2)$.

The size of the conditioning set is driven by the relevance decay incorporated into the prior for $f_n$. This decay follows an exponential pattern based on the position within the conditioning set. Consequently, we limit the conditioning set size such that the conditioning set only comprises neighbors whose contributions are still considered relevant, e.g., $m = \max \{j \geq 1 : \exp(-j \exp(\theta^q)) \geq \epsilon \}$. In our analysis, we set $\epsilon = 0.01$ and estimate the value of $\theta^q$. Thus, the size of the conditioning set is automatically determined.

\citet{Katzfuss2021} suggest employing an empirical Bayes (EB) approach for inference. Due to conjugacy, the integrated likelihood (with $\bf$ and $\bd$ integrated out) in Equation~\eqref{eq:lik-fact} is available in closed form and can be maximized using numerical methods, leading to $\hat\bftheta = \argmax_\theta p(\bY | \bftheta)$. For the spatial case, the authors discuss details that motivate the prior choice, the relationship between $\bftheta$ and $\bfpsi$, as well as closed-form expressions for the integrated likelihood and predictive posterior distribution of $\by | \bY, \hat\bftheta$.

\subsection{Extension to multivariate spatial fields}

As in the univariate case, we aim to estimate the joint distribution of $\bm{y}$, allowing for non-Gaussian dependence.
We propose employing the BTM approach in a higher-dimensional input space $\tilde{\mathcal{S}}$.
We construct the new input space by combining the spatial domain $\mathcal{S}$ with a latent process space $\breve{\mathcal{S}} = \mathds{R}^{P-1}$ in which the marginal spatial fields are positioned relative to each other.
These process positions $\breve{\bs}_p \in \breve{\mathcal{S}}, p=1, \dots, P$, serve to position the univariate spatial fields relative to one another.
In the latent process space, smaller distances between $\breve{\bs}_p$ and $\breve{\bs}_{p'}$ indicate stronger associations between fields $\by_p$ and $\by_{p'}$.

\begin{figure}[bt]
    \centering
     \includegraphics[width=\textwidth]{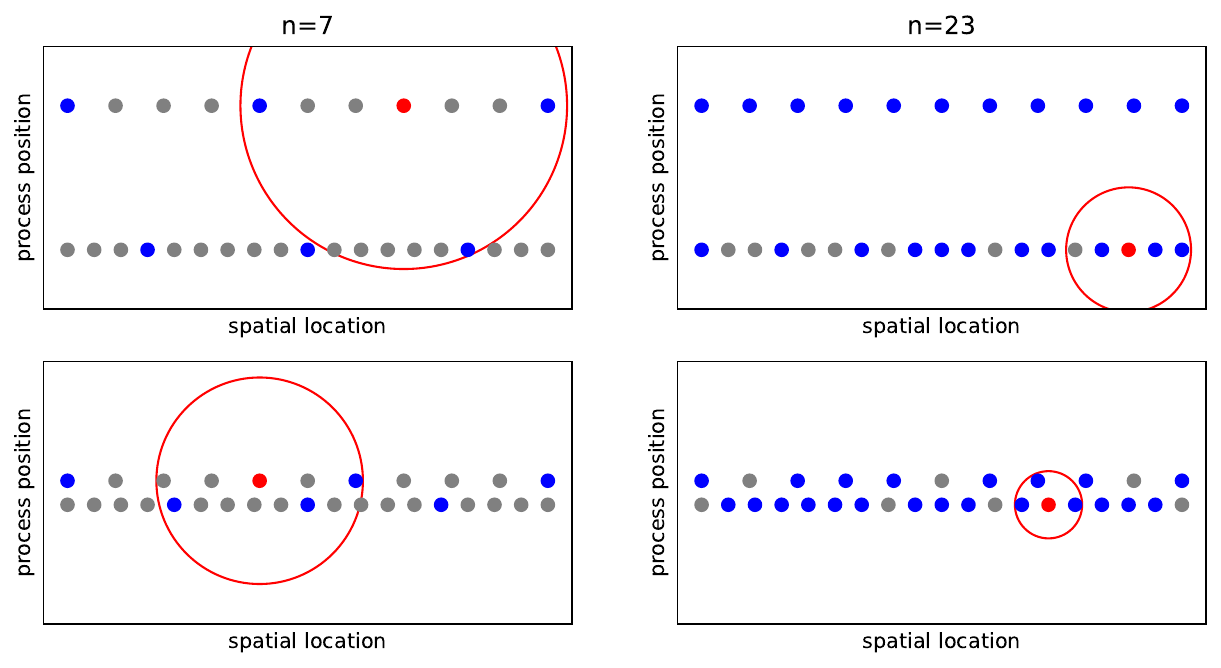}
    \caption{
    Stylized illustration of the augmented input space for two one-dimensional spatial fields ($P=2$ and $\operatorname{dim}(\mathcal{S}) = 1$).
    The top row corresponds to weakly associated fields, while the bottom row represents strongly associated fields.
	The dots in the figure represent locations within the augmented input space.
	The colors of the dot correspond to the position relative to $n$ in the maxmin ordering. The $n$-th ordered location is colored red, while the locations previously ordered are colored blue and the locations subsequent in the ordering are colored gray.
	Here, the situation for $n$ equals 7 (left column) and $n$ equals 23 (right column) are considered.
	The circles in the figure depict the radius of the largest distance from the $n$-th location to locations included in the reduced conditioning set $c(n)$ of size $m=3$.
 The conditioning sets are constructed using the three nearest neighbors.
	It is important to note that only observations preceding the $n$-th location in the maxmin ordering, these are the blue colored dots, can be included in the conditioning set.
	Consequently, blue dots within the circle form the conditioning set for the red dot.
	Analyzing the illustration, we can observe that the conditioning set comprises observations from both fields in the early stages of the ordering.
	However, as the index progresses in the  ordering, the conditioning set primarily includes spatial neighbors from the same field.
	This effect is more pronounced for weakly associated fields.
    }
	\label{fig:illustration-inputspace}
\end{figure}

We combine the process space with the spatial domain, resulting in the augmented input space $\tilde{\mathcal{S}} = \mathcal{S} \times \breve{\mathcal{S}}$.
Now, each element $y_n$ of $\by$ is associated with a location in the augmented input space, represented by $\tilde \bs_n \in \tilde{\mathcal{S}}$.
Figure~\ref{fig:illustration-inputspace} provides a visual representation of this concept using a toy example consisting of a two-variate field on a one-dimensional spatial domain.
The figure displays the spatial domain along the x-axis and the process space along the y-axis. Each point in the plot represents a location in the augmented input space, combining the spatial domain and the process space.
The top row of the figure depicts less-correlated processes with a greater distance between them, while the lower row illustrates strongly dependent processes with a smaller distance. 
In the figure, two situations for the position in the ordering are considered, one early ($n=7$) and one late ($n=23$) in the ordering. The locations are ordered according to maxmin ordering and the reduced conditioning sets are indicated by the red circles, showcasing how reduced conditioning sets early in the ordering consist of observations from both processes while later in the ordering observations from the other process enter the conditioning sets only if the processes are close. The latter part is explained with more detail below.

\subsubsection{Vecchia approximation in the augmented input space}\label{sec:meth-vecchia}

The transport map formulation casts the problem of inferring the distribution of $\by$ into the task of solving $N$ independent regression problems of the form
\begin{align}
	y_n |f_n, d_n, \by_{1:n-1} &\sim \normal(f_n(\by_{1:n-1}), d_n^2)\\
	f_n | d_n, K_n & \sim \GP(0, d_n^2 K_n)\\
	d^2_n | \alpha_n, \beta_n &\sim \mathcal{IG}(\alpha_n, \beta_n).
\end{align}
As described above, the estimation is computationally infeasible for larger dataset sets.
Similar to \citet{Katzfuss2021}, we address the problem by conditioning $y_n$ only on the sub-vector $\by_{c(n)}$, $c(n) \subseteq \{1, \dots, n-1\}$ of $\by_{1:n-1}$, rendering $y_n$ independent of $\by_{\{1, \dots, n - 1\} \setminus c(n)}$ given $\by_{c(n)}$, $f_n$, and $d_n$.

Our method uses the Euclidean distance within the augmented input space for computing the maxmin ordering and the conditioning sets.
Following the maxmin ordering, we define the $n$-th index in the sequence $\Omega$ that orders $\by$ such that it maximizes the minimum distance to all previously ordered locations. In mathematical terms,
\begin{align*}
\Omega_n = \argmax_{i \in \{1, \dots, N\} \setminus \Omega_{1:n-1}} \min_{j \in \Omega_{1:n-1}} \Vert \tilde{\bs}_i -  \tilde{\bs}_j \Vert.
\end{align*}
Notably, the first index in $\Omega$ can be selected arbitrarily, and in our implementation, we opt for the most central point as the initial choice.
This ordering concept exhibits parallels with the space-filling maxmin-distance design \citep{johnson1990minimax,pronzatoDesignComputerExperiments2012}, although it operates within the spatial constraints of available locations. Furthermore, it applies the maxmin criterion sequentially in each step, as opposed to global optimization.
We view this ordering as space-filling on different resolutions, initially filling the space on a coarse scale and subsequently decreasing the distance to the nearest location.

Following the literature, we define reduced conditioning sets $c(n)$ as the $\min(n - 1, m)$ nearest (with respect to Euclidean distance) previously ordered neighbors. More precisely, the $i$-th element is given as
$c(n)_i = \argmin_{j \in \{1, \dots, n-1\} \setminus c(n)_{1:i-1}} \Vert \tilde{\bs}_n - \tilde{\bs}_j \Vert$.
This choice is motivated by the so-called screening effect \citep{Stein2011}, which states that for many popular covariance functions, such as the Matérn covariance function, random variables in a spatial field are (almost) independent of distant random variables conditioned on the values in-between.

In a stylized example shown in Figure~\ref{fig:illustration-inputspace}, we demonstrate this approach.
Consider two relatively smooth spatial fields with a weak association between them.
In this scenario, the distance to the nearest spatial neighbor is smaller than the distance to the next process.
As we employ the maxmin ordering, locations chosen almost alternate between processes early in the ordering.
Consequently, the first conditioning sets are likely to contain information from both processes.
However, as the ordering progresses, the conditioning sets will predominantly include information about the spatial neighbors.
On the other hand, when dealing with fields that exhibit a strong association, the conditioning sets will encompass a larger number of values from the other process even later in the ordering.

Using this Vecchia-type approximation introduces dependence of $\by$'s modeled distribution on its order, and the choice of the conditioning sets $c_{(n)}$.
The combination of employing the maxmin ordering and selecting nearest neighbors for the conditioning set has been widely recognized in the literature as an effective approach \citep{Datta2016, Heaton2017, Huang2021, Katzfuss2017a}.

\subsubsection{Distance metric}

Besides their direct involvement in the prior construction, the input locations, or, more strictly speaking, the distances, also determine the ordering and are used to find the nearest neighbors. Both are essential to the quality of the Vecchia approximations. The involvement of distances in the posterior density allows gradient computations for the distances and, consequently, the process positions, thus, enabling a gradient-based update of the process positions.

Conversely, choosing an adequate distance metric is crucial for the model's quality.
Using the Euclidean distance may be a natural choice when applied to spatial locations.
In other scenarios, including the multivariate case discussed in this article, the Euclidean distance is not necessarily meaningful, and other distance measures can be used.
For example, \citet{Kang2021} explore a correlation-based distance metric for GP regression.
In contrast, we propose employing the Euclidean distance in the augmented input space.
We argue that the Euclidean distance is meaningful in the higher-dimensional augmented input space since the distance between each input pair $\tilde{\bs}_p, \tilde{\bs}_{p'}$ is composed of both the distance between the spatial locations and the distance between the process locations.
By appropriately scaling the latent process space, the distances between the processes can be interpreted similarly to distances in the spatial domain.

\subsubsection{Parameterization of the process positions}

We aim to include the process positions in the hyperparameter vector $\bftheta$.
However, the model depends only on the relative process positions, because the model depends on the input locations only through the distances.
For any two indices $n$ and $n'$ within the range of ${1, \dots, N}$, the squared Euclidean distance between the augmented input locations $\tilde\bs_{n}$ and $\tilde\bs_{n'}$ can be decomposed as $\Vert \tilde\bs_{n} - \tilde\bs_{n'}\Vert^2 = \Vert \bs_{n} - \bs_{n'}\Vert^2 + \Vert \breve\bs_{n} - \breve\bs_{n'}\Vert^2$.
Here, $\bs_{n}$ represents the spatial location and $\breve\bs_{n}$ represents the process position associated with $y_n$; analogous for symbols with index $n'$.
As the model depends on the process positions only through their distances, we can fix rotation and the first location.
For this purpose, we define $\breve{\bs}_1 = \bm{0}$ and decrease the degrees of freedom for each subsequent process position from $P-1$ to $1$ by one at a time.
Jointly, the process positions $\breve{\bS}$ are parameterized as 
\begin{equation}
	\breve{\bS} = \begin{pmatrix} \breve{\bs}_1^T \\ \vdots \\ \breve{\bs}_P^T\end{pmatrix} = \begin{pmatrix} \bm{0}^T \\ \bQ \bR\end{pmatrix},
\end{equation}
where the columns of $\bQ$ form an orthonormal basis of $\mathds{R}^{P-1}$ and $\bR$ is an upper triangular matrix with positive entries on the diagonal.
Only the $P(P-1)/2$ nonzero entries of $\bR$ are included as hyperparameters in the model.

\subsection{Estimation}

The estimation procedure consists of two stages.
In the first stage, we fit a parametric and separable GP model to obtain initial values for the process positions $\breve{\bS}$.
These initial values serve as a starting point for the subsequent estimation of the transport map hyperparameters.
For the estimation of the hyperparameters, we explore three different approaches, which are detailed below.
Finally, using the estimated hyperparameters, the transport map can be employed as a generative model to simulate new samples that capture the spatial dependencies as learned from the data.
The transport map can also be used for uncertainty quantification.
The multivariate extension allows us to study the conditional distribution of the spatial fields corresponding to one or more variables given the observed spatial fields of other variables.

 \subsubsection{Obtaining the initial process positions}\label{sec:meth-initial-process}
 
 To determine the initial process positions, specifically $\breve{\bs}_j \in \mathcal{S}_p$ referring to the second part of the augmented input locations, we fit a separable parametric model.
  In this model, we evaluate a separable covariance function parameterized with $\bfzeta$ at the locations $\bs_1, \dots, \bs_N$ to obtain the covariance matrix $\bm{K}_{\bfzeta}$.
  For the two elements  $y_{n}, y_{n'}$ with the associated spatial locations $\bs_{n}, \bs_{n'}$ and processes $p, p'$, the covariance between $y_n$ and  $y_{n'}$ is given by
  $$\mathrm{Cov}[y_n, y_{n'} | \bfzeta] = \tau^2_{\bfzeta} C_1(\Vert \bs_{n} - \bs_{n'}\Vert | \bfzeta) C_2(p, p' | \bfzeta) + \sigma^2_{\bfzeta}\mathds{1}_{\bs_n = \bs_n' \wedge p = p'}.$$
 Here, $\tau^2_{\bfzeta}$, $\sigma^2_{\bfzeta}$ is are variance parameters, $C_1$ corresponds to a parametric isotropic correlation function, and $C_2$ corresponds to an unstructured $P \times P$ correlation matrix $\breve{\bK}_{\bfzeta}$.
Moreover, $\mathds{1}$ is the indicator function.
To map $\binom{P}{2}$ unrestricted elements to $\breve{\bK}_{\bfzeta}$, we employ the mapping described in \citet[Section 10.12]{stan2023}.

To estimate $\bfzeta$, we maximize the likelihood of $\by \sim \normal(\bm{0}, \bm{K}_{\bfzeta})$ by considering all available samples of $\by$.
Considering the availability of multiple observations of $\by$, we have found that thinning the spatial density can effectively manage computational time constraints.
By randomly choosing a subset of locations, we can alleviate the computational burden associated with factorizing the covariance matrix $\bm{K}_{\bfzeta}$.
This approach strikes a balance between computational efficiency and ensuring reliable parameter estimation for $\bfzeta$.

\paragraph{Retrieving the hyperparameter values}
Using the estimated parameter $\hat\bfzeta$, a distance between two marginal spatial fields can be computed by applying the inverse of the correlation function $C_1$ on the value of $C_2$.
For instance, the estimated distance between the $p$-th and $p'$-th marginal spatial fields is $\hat d_{pp'} = C_1^{-1}(|C_2(p, p' | \hat{\bfzeta})| | \hat{\bfzeta})$.
Arranging the estimated distances in a $P \times P$ distance matrix $\hat\bD = (\hat d_{ij})_{1 \leq i \leq P, 1 \leq j \leq P}$ allows us to easily compute a set of process positions that give rise to $\hat \bD$ \citep{youngDiscussionSetPoints1938, torgersonMultidimensionalScalingTheory1952}.

Given a distance matrix $\bD = (d_{ij})_{1 \leq i \leq P, 1 \leq j \leq P}$, it is possible to determine a set of locations that generate this matrix up to Euclidean transformations.
Let the elements of the matrix $\bE = (e_{ij})_{1 \leq i \leq P, 1 \leq j \leq P}$ be $e_{ij} = d_{1, j}^2 + d_{i0}^2 - d_{ij}^2$.
Then, the matrix representing the process locations is $\breve \bS = \bU \bfLambda^{1/2}$ where $\bU$ and $\bfLambda$ are obtained by performing the eigenvalue decomposition $\bE = \bU \bfLambda \bU^{-1}$ and defining $\bfLambda^{1/2}$ as the element-wise square root.
Subsequently, a QR decomposition can be performed on the bottom $P-1$ rows of $\breve \bS$ to calculate $\bQ$ and $\bR$.
Should the values on the diagonal of $\bR$ not be positive, use $\bQ\bm{\Upsilon}$ and $\bm{\Upsilon}\bR$ instead of $\bQ$ and $\bR$ where $\bm{\Upsilon}$ is a square diagonal matrix with the entries 1 or -1 corresponding to the sign of the diagonal elements of $\bR$.

\subsubsection{Estimation of the multivariate transport map}\label{sec:estimation-tm}

We extend the empirical Bayes approach used by \citet{Katzfuss2021} to estimate the hyperparameters $\bftheta$.
The non-zero elements of $\bR$ are included in $\bftheta$, and suitable transformations are applied to ensure that $\bftheta$ remains unrestricted.

To estimate the hyperparameters $\bftheta$, we employ a gradient-based optimization algorithm to maximize the integrated likelihood
\begin{align}
	\label{eq:lik-fact2}
	p(\by | \bftheta) = \prod_{n=1}^{N} \int p(y_n | \by_{c(n)}, d_n, f_n) p(f_n, d_n | \bftheta)\;\text{d}{d_n}\;\text{d}f_n. 
\end{align}
The presented method focuses on inferring the distribution of $\by$ rather than on conducting inference for $\bf$ and $\bd$, providing a significant computational advantage, especially since the formulation with $\bf$ and $\bd$ integrated out permits the utilization of mini-batching, allowing for efficient computations on data subsets.
We utilize the Adam optimizer \citep{kingma2014adam}, which adapts the learning rate during optimization, and implement early stopping as a regularization technique.
The early stopping criterion is based on monitoring the improvement of the integrated log-likelihood on a separate validation dataset.
We define a patience parameter, typically set to 5\% or 10\% of the maximum number of iterations.
If no improvement in the integrated log-likelihood is observed within the last patience steps, the optimization process is terminated.
Upon completion of the optimization, we select the parameter set that achieved the highest integrated log-likelihood on the test data.
This parameter set represents the optimal configuration based on the performance of the model on unseen data.

However, in terms of estimating the process positions, we consider three strategies:

\begin{enumerate}
	\item[i)] Constant process positions (CPP): In this strategy, the process positions, estimated using the parametric model, are assumed to be known and remain fixed during the hyperparameter estimation process. No updates to the process positions are made.

	\item[ii)] Frozen Ordering (FO): With this approach, the process positions are updated during the hyperparameter estimation via a gradient-based update. Recall, the process positions enter the integrated likelihood through the integration of the distance to the nearest neighbor $\ell_n$ in the prior on $d_n^2$.
	The ordering and conditioning sets, however, are determined using the initial process positions and remain unchanged throughout the optimization.

	\item[iii)] Occasional Re-ordering (OR): Here, the process positions are updated during the hyperparameter estimation via a gradient-based update. Following the idea of \citet{Kang2021}, we recompute the ordering and conditioning sets after a pre-specified number of iterations (e.g., after $4, 8, 16, 32, \dots$ iterations). 
	
	  As the integrated likelihood is based on a Vecchia approximation (using Equation~\eqref{eq:lik-fact2} instead of Equation~\eqref{eq:lik-fact}), reordering may produce distinct likelihood values. Furthermore, the reordering procedure does not necessarily lead to an improvement in the integrated likelihood. 
	Hence, we reset the patience counter and update the best encountered integrated likelihood to the current value. This ensures that the algorithm will continue in searching for a better parameter value for at least the specified number of epochs and considers only the integrated likelihood values encountered in the current ordering.
	  Additionally, determining these updates can be computationally expensive as one has to consider the entire data set. However, \citet{Schafer2017} presented an algorithm to compute ordering and conditioning sets in quasi-linear time complexity, i.e., $\mathcal{O}(N \log^2(N))$.
\end{enumerate}
For ease of comprehension, we provide the estimation algorithm in pseudo-code in Appendix~\ref{app:algorithm}.

\section{Numerical comparison \label{sec:numerical}}

\subsection{Simulation Study}\label{sec:sim-study}

Our study focuses on two objectives: (1) learning the distribution of $\by$ and (2) learning the conditional distribution of one variable given the others, i.e., the distribution of $\by^{1} | \by^{2}, \dots, \by^{P}$. Additionally, we aim to compare the performance of the different estimation strategies: Constant process positions (CPP), Frozen Ordering (FO), and Occasional Re-ordering (OR).
To assess the of the MVTM, we compare it with a parametric model.
We refrain from a comparison to VAE \citep{Kingma2014} and a GAN designed for climate-model output \citep{Besombes2021} as those deep-learning methods have turned out as not competitive in an application similar to ours \cite[see the supplementary materials in][for details]{Katzfuss2021}.

To evaluate the performance, we assess the average log-density of the learned distribution at 20~test samples.
This evaluation metric provides an approximation, up to an additive constant, of the negative Kullback-Leibler (KL) divergence $D(p || \hat{p})$ between the true distribution $p$ and the estimated distribution $\hat{p}$.
The log-density serves as a positively oriented measure, allowing us to compare the accuracy and goodness-of-fit of the learned distribution to the true distribution.

\subsubsection{Experimental Setup}

We vary two key factors in our simulation study: the training size and the number of variables. Specifically, we consider training sizes $R \in \{10, 20, 30, 40, 60, 80\}$, and explore scenarios with $P \in \{2,3,4,5\}$ variables. By varying these factors, we aim to examine the behavior and performance of the estimation procedures under different data settings.

For the data-generating process, we adopt Scenario NR900 in Katzfuss et al. (2021). This data-generating process can be characterized by a transport map. Using the hierarchical formulation in Section~\ref{sec:meth-vecchia}, we specify the functions $f_i$ as additively composed from a linear and non-linear part
$$
f_i(\by_{1:i-1}) = \bb_{i}^T\by_{c(i)} + 2\sin(4(b_{i,1} y_{c(i)_1} + b_{i,2} y_{c(i)_2}))
$$
with $\bb_i = (b_{i,1}, \dots, b_{i, |c(i)|})^T$, where the $b_{i,k}$ are based on the exponential covariance function with range $0.3$ and the distance between $\tilde{\bs}_i$ and $\tilde{\bs}_{c(i)_k}$.
The $P$-variate spatial field is observed on a regular grid of size $32\times 32$ on the unit-square and we use the augmented input space with the process positions $\breve{\bs}_1 = \bm{0}$, $\breve{\bs}_2 = (0.2, 0, 0)^T$, $\breve{\bs}_3 = (0.0, 0.3, 0)^T$, $\breve{\bs}_4 = (0.0, 0, 0.4)^T$, and $\breve{\bs}_5 = (0.3, 0.3, 0)^T$.
Thus, the $P$-variate spatial field is observed in 1,024~locations per process, which gives 5,120 locations for a five-variate spatial field in the augmented input space.
In total, roughly 400,000~datapoints must be considered in the scenario with 80~replicates of the five-variate field.

For fast estimation of the $6 + \binom{P}{2}$ hyperparameters $\bftheta$, we utilize mini-batching with a batch size of 256, resulting in $4P$ gradient updates per epoch.
Convergence monitoring is conducted by evaluating the integrated log-likelihood on an independently generated validation dataset with 20~replicates.
We employ early stopping with a patience of 25 and a maximum of 500 iterations.
Remarkably, all estimations terminate early, indicating successful convergence.
To optimize the model, we utilize the Adam optimizer with an initial learning rate of 0.01 and apply cosine annealing to mitigate the variance induced by mini-batching.
For estimating the initial process positions, we randomly select 256~of the 1,024~spatial locations.
Importantly, we observe no adverse effects on the estimated parameters due to the subsampling.
In our evaluation, we also include as a competitor procedure a mean zero Gaussian model with isotropic Matérn covariance function combined with an unstructured $P \times P$ correlation matrix as described in Section~\ref{sec:meth-initial-process}, whose ${3 + \binom{P}{2}}$~hyperparameters are estimated via maximum likelihood. We refer to this model as the parametric model.

\subsubsection{Results and Analysis}

\begin{figure}[t]
    \centering
    \includegraphics[width=\textwidth]{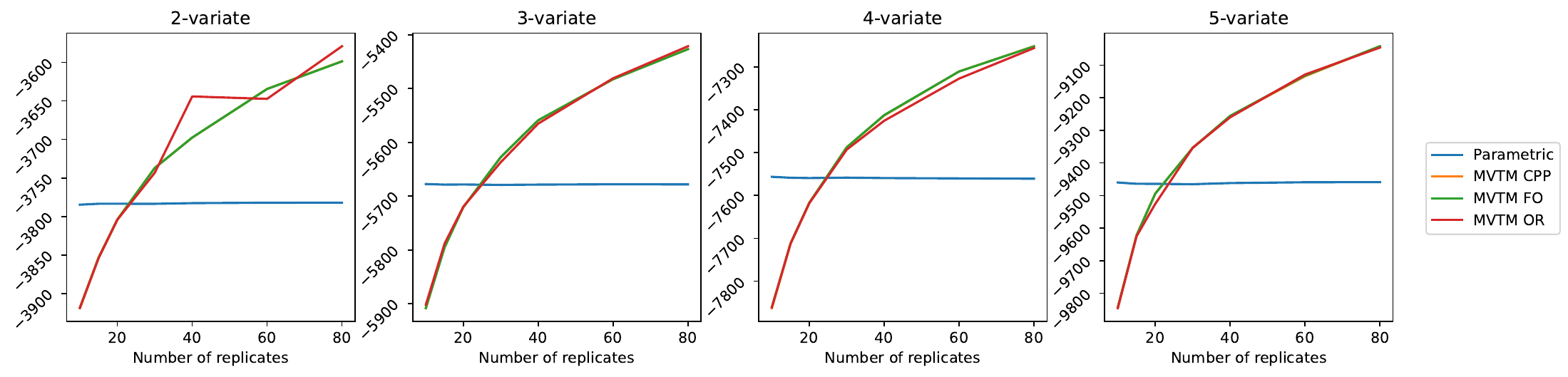}
    \caption{
        Mean of the estimated log-density of test data sets for different ensemble sizes $R$ (x-axis) and varying dimension $P$ of the multivariate spatial field (panels). The ``Parametric'' approach refers to the parametric GP, while approaches labelled with ``MVTM CPP'', ``MVTM FO'', and ``MVTM OR'' correspond to the multivariate Bayesian transport map with the estimation strategies described in Section~\ref{sec:estimation-tm}. Note that the orange line is almost completely covered by the green line.
    }
	\label{fig:sim-log-scores}
\end{figure}

The results indicate that the transport map outperforms the parametric model when the ensemble size exceeds approximately 20 to 25 (refer to Figure~\ref{fig:sim-log-scores}).
Interestingly, incorporating the process positions into the hyperparameter estimation of the transport map (OR) did not lead to an improvement in model fit.
However, it is worth noting that this might be attributed to the limited nonlinearity in the data-generating process.

\begin{figure}[b]
    \centering
    \includegraphics[width=\textwidth]{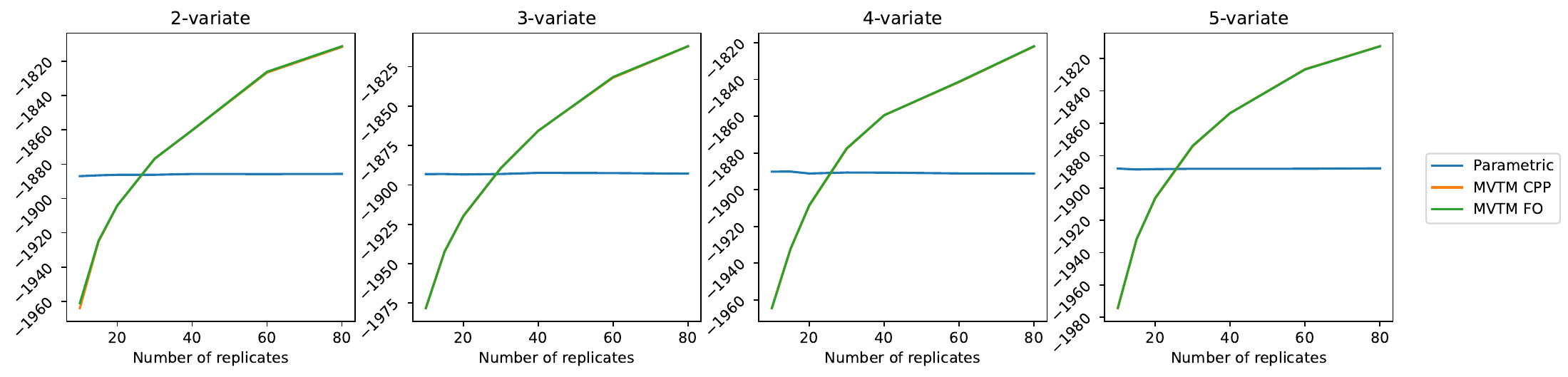}
    \caption{Log-density of $\by_1$ conditioned on $\by_2, \dots, \by_P$. The plots show the mean log-density evaluated on test data sets for different ensemble sizes $R$ (x-axis) and different dimensions $P$ of the multivariate spatial field.  ``Parametric'' refers to the parametric approach, ``MVTM CPP'' and ``MVTM FO'' refer to the Bayesian transport map with the estimation strategies described in Section~\ref{sec:estimation-tm}.
    }
    \label{fig:sim-log-scores-cond}
\end{figure}

In addition to estimating the joint distribution of $\by$, we also investigate the conditional distribution of $\by_1$ given $\by_2, \dots, \by_P$. To account for the conditional distribution, we use a modified maxmin-ordering in which we order the indices corresponding to $\by_1$ subsequent to all other indices. Based on the findings from the initial study, we exclude the OR estimation strategy from this analysis as it did not yield improved results. Again, the approximated KL-divergences suggest that the MVTM is superior given enough training data. See Figure~\ref{fig:sim-log-scores-cond} for a visual presentation of the estimated log densities.

In summary, the findings from our simulation study indicate that the MVTM approach performs better when a sufficient amount of training data is available.
We observe that the MVTM approach outperforms the alternative methods in capturing the underlying spatial and inter-process dependencies given enough available training data.
Interestingly, the choice of estimation strategies seems to have no significant impact on the performance.
This lack of impact could be due to the limited non-linearity present in the data generating process.
Specifically, only the two nearest neighbors in the conditioning set, which belong to the same process for a large portion of the data, contribute to the conditional expected value in a non-linear manner.
Consequently, the parametric approach is able to effectively capture the dependence between the spatial fields.
We also explored an alternative configuration, involving the scaling of weights $b_{i,1}$ and $b_{i,2}$ based on the process affiliations of both $y_i$ and its corresponding neighbors $y_{c(i)_1}$ and $y_{c(i)_2}$. This leads to a scaling of the non-linear functions based on process affiliations. We omit a detailed presentation, as the results are comparable, with MVTM outperforming the parametric GP with only $R=20$ training replicates.
Note that our study does not explore spatial data sparsity or density, as previous results by \citet{Katzfuss2021} suggests that the performance differences among the methods are similar under such conditions.
Therefore, in the context of our simulation study, the focus is primarily on the performance of the MVTM approach in relation to the availability of training data.

\subsection{Application: Climate model output}\label{sec:application}

Climate models serve as computational tools essential for simulating and
comprehending the Earth's climate system, playing a vital role in climate
research and policy-making. In essence, climate models are computer programs that
describe the Earth's climate system through sets of differential equations.
Developing as well as running these models demands a lot of resources, time, and specialized high-performance computers.
Each run takes a considerable amount of time and consumes significant energy.
For instance, 17~million core hours were spent on the computation of large ensemble of the Community Earth System Model~(CESM) using the Yellowstone supercomputer \citep{kay2015CESM}.
It takes approximately three weeks to produce each ensemble member.

Nevertheless, conducting multiple runs is typically imperative due to the
potential impact of even slight perturbations in the initial conditions, which
can result in larger variations at the end of a model run.
Therefore, it is crucial to perform multiple runs of climate models to account for the uncertainty arising from these initial conditions.
Due to this nature, a climate model can be interpreted as encoding a distribution of climate rather than predicting an exact Earth-system state, rendering the analysis of climate model even more challenging.
Statistical emulators, which replicate this distribution, can be employed to summarize the distribution and generate additional samples at much lower computational costs.

Climate models produce numerous variables with high temporal and spatial resolution, often exhibiting significant non-stationarity.
In this study, we specifically examine data obtained from the large ensemble project (LENS) of the CESM developed by the National Center for Atmospheric Research.
The LENS consists of 42~ensemble members with variations due to slightly different initial atmospheric state.

Our focus centers on four hydrological variables generated by the land-surface-model (LSM) component\footnote{Data is available at \url{https://www.earthsystemgrid.org/dataset/ucar.cgd.ccsm4.cesmLE.html}.}.
We restrict our focus further to the conterminous United States and  consider the yearly average of the variables \emph{SNOW}, \emph{RAIN}, \emph{SOILWATER\_10CM}, and \emph{QRUNOFF} in the year 2001.
The data is available on a roughly $1^\circ$~longitude-latitude grid yielding 690~values per variable and ensemble member.
In a prepossessing step, we combine the first two variables to a new variable representing the combined precipitation (\emph{PRECIP}).
As we aim to model anomalies, all raw observations are transformed by subtracting the pixel-wise mean and dividing by the pixel-wise standard deviation.
\begin{figure}[tb]
    \centering
    \includegraphics[width=\textwidth]{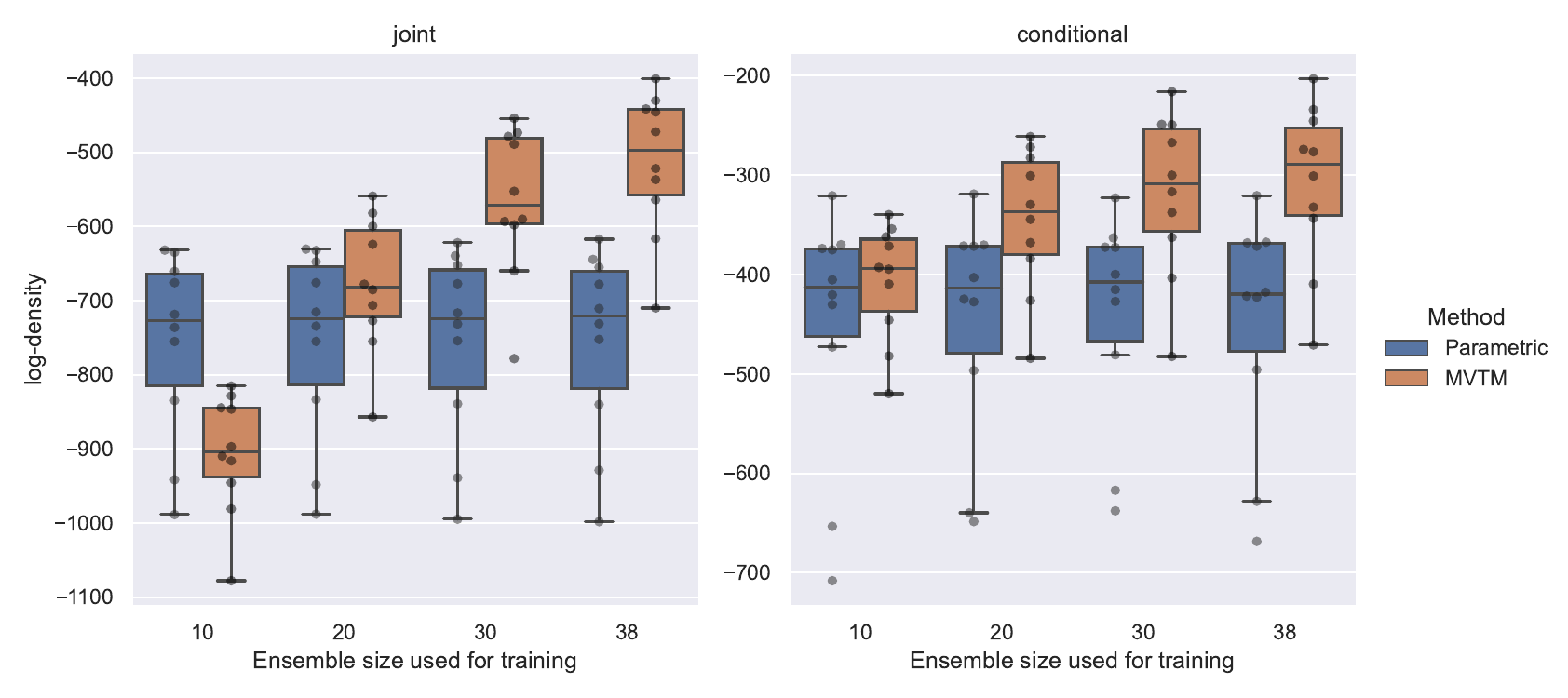}
    \caption{
        The box plots illustrate the estimated log-density of the holdout data, with the points representing the underlying data. The left panel concerns the log-density of the joint model while the right panel depicts the log-density of \emph{QRUNOFF} conditioned on the remaining variables. The x-axis indicates the size of the ensemble used for training.
    }
    \label{fig:appl-log-scores}
\end{figure}

\begin{figure}[b]
    \centering
    \includegraphics[width=\textwidth]{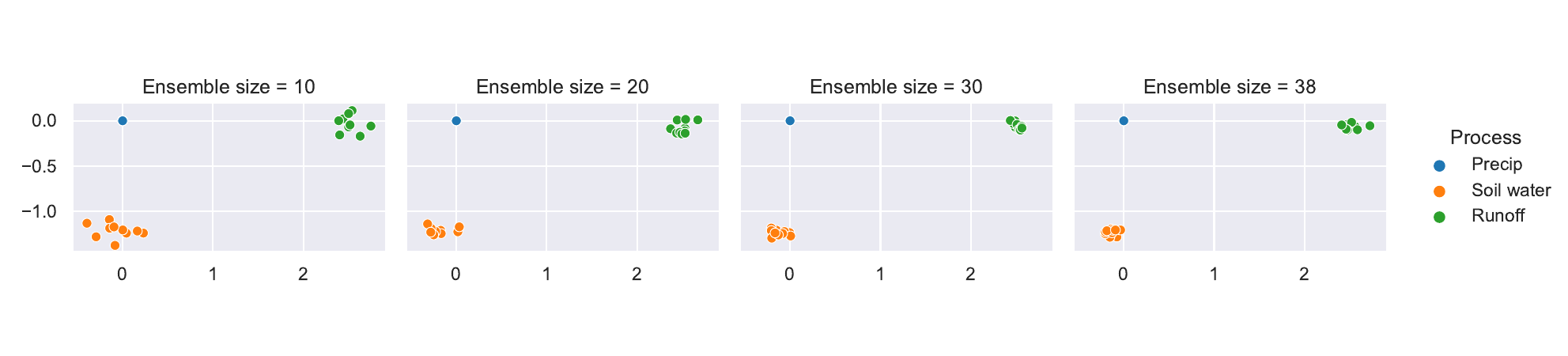}
    \caption{
        Scatterplots of the latent positions estimated in the joined model for 10 CV splits, stratified by ensemble size used for training. Each point represents the estimated position of process in one replication. The location of the first process (PRECIP) is by construction $\bm{0}$.
    }
    \label{fig:appl-latent-pos}
\end{figure}

We fit the MVTM using estimation strategy CPP and compare it to the parametric model employing the log-density evaluated for holdout data.
Since the other estimation strategies produce similar results, we omit their presentation.
In a cross-validation (CV) setting, we select 4~ensemble members as holdout data.
From the remaining 38 members, we use 10, 20, 30 and 38~for training to investigate the effect of sample size.
The process is repeated until all but the last two data sets are once used as holdout, giving us ten values for the estimated log-density.
The left panel in Figure~\ref{fig:appl-log-scores} visualizes the results, strengthening the impression from the simulation study.
The MVTM's performance improves with increasing samples size.
In comparison, it adapts better to the underlying dependence in the anomalies.
The estimated latent locations are illustrated in Figure~\ref{fig:appl-latent-pos}.
Notably, these estimated locations exhibit relatively little variation across CV splits as the training size increases. 
Nevertheless, it is worth noting a discernible trend of increased point concentration among replications as the sample size grows.

On an Apple M1 Pro equipped with eight performance cores and 32GB of RAM, the estimation process takes an average of 91 seconds of CPU time for ten replications and 194 seconds for 38~replications. This estimation time is distributed between two tasks: estimating the initial latent locations and the hyperparameters of the transport map. Specifically, estimating the initial latent locations consumes 12, 22, 33, and 44~seconds for 10, 20, 30, and 38~replications, respectively. Meanwhile, the estimation of hyperparameters requires between 79, 98, 122, and 150~seconds for 10, 20, 30, and 38~replications, respectively. The standard deviation of these estimations is roughly 1\%, indicating a relatively stable performance across CV-replications.

\begin{figure}
\centering
\begin{subfigure}{.45\textwidth}
  \centering
  \includegraphics[width=2.7\linewidth, angle=270, origin=c]{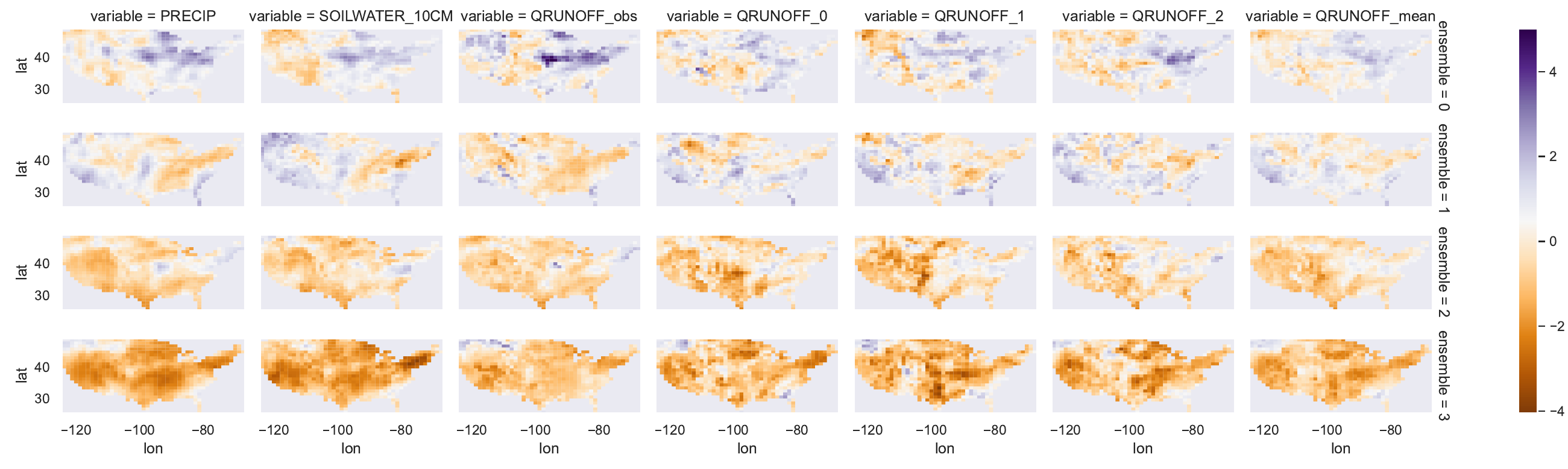}
  \caption{Model: MVTM}
  \label{fig:sub1}
\end{subfigure}%
\begin{subfigure}{.45\textwidth}
  \centering
  \includegraphics[width=2.7\linewidth, angle=270, origin=c]{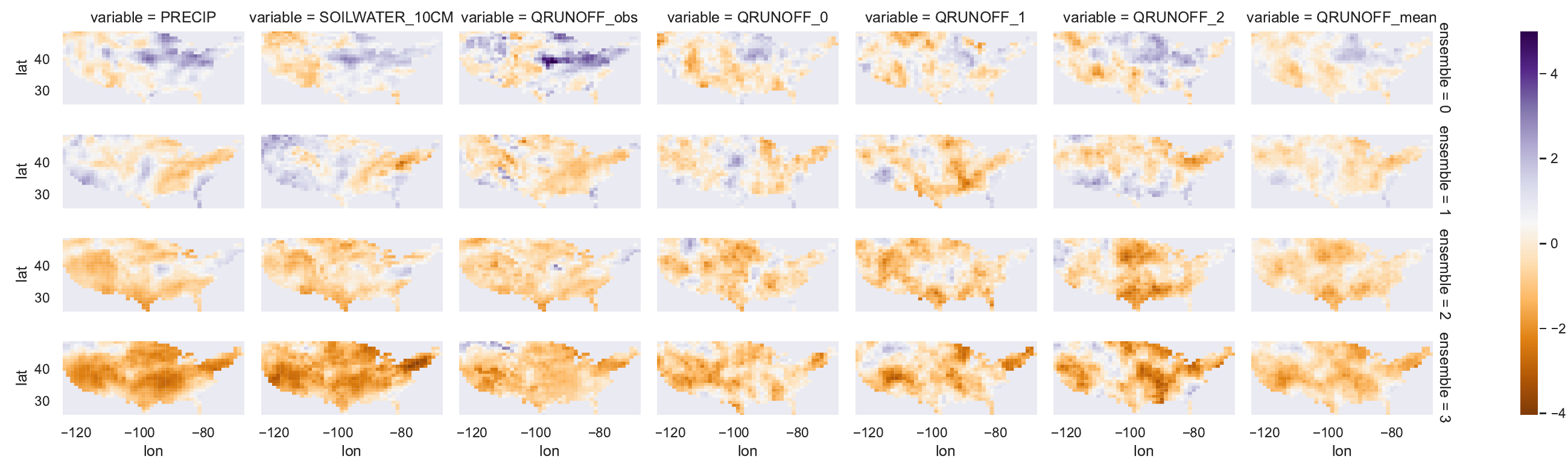}
  \caption{Model: Parametric}
  \label{fig:sub2}
\end{subfigure}
\caption{
Plots of the anomalies of the variables PRECIP, SOILWATER\_10CM, and QRUNOFF. The other columns show samples of QRUNOFF drawn from the modeled distribution when conditioning on PRECIP and SOILWATER\_10CM. The last column shows the pixel-wise mean based on 30~samples. The rows correspond to different ensemble members. Panel (a) is generated using the MVTM and Panel (b) using the separable parametric model.
}
\label{fig:cond_samples}
\end{figure}

In addition to studying the joint distribution of the variables, we also investigate the conditional distribution of \emph{QRUNOFF} given the other two variables.
This analysis is particularly valuable as runoff in an area cannot be directly measured but must be modeled. Having a statistical model that describes the conditional distribution of runoff can be immensely beneficial for practitioners.
For example, it enables the study of the conditional distribution of catchment runoff within the drainage basin of a river, which represents the water from sources like rain, snow, and soil moisture flowing into the river.
Thus, a model that characterizes the conditional distribution of runoff in an area based on observable quantities can assist in estimating the amount of water a river system needs to handle and evaluating whether infrastructure, such as dams, is adequately designed.
From the right panel in Figure~\ref{fig:appl-log-scores}, we see that the MVTM describes the conditional distribution already for the smallest training size better than the parametric model.
Judging from visual impression of samples drawn from the estimated distribution, the parametric model cannot capture the anisotropy present in the data and seems overall too smooth (see Figure~\ref{fig:cond_samples}).

\section{Conclusions \label{sec:conclusions}}

We have presented a Bayesian approach for learning the distribution of multivariate spatial fields based on a relatively small number of training samples by estimating a transport map.
Our method can capture the potential non-linearity in the conditional dependencies, enabling the learning of non-Gaussian distributions and mapping them to the standard normal.
The scalability of the approach is enhanced by the mini-batching capability of the estimation algorithm, allowing for efficient analysis of large datasets.
From our numerical demonstrations, we can confidently recommend the multivariate transport map  approach, particularly when dealing with non-stationarity in the data and an ample number of replications.

We do not account for uncertainty in the estimation of the hyperparameters $\bftheta$. While fully Bayesian approaches like Markov chain Monte Carlo (MCMC) can in principle address this, we opt for using empirical Bayes (EB) estimation due to computational constraints. This decision is supported by insights from \citet{Katzfuss2021}, which indicate minimal impact on the estimated posterior distribution of $\by$ when contrasting EB results with a full Bayesian approach. MCMC's requirement for full dataset processing in each iteration poses scalability challenges for large datasets. Stochastic gradient MCMC methods \citep{nemeth2021stochastic} show promise in bypassing this limitation, offering potential for enhanced hyperparameter uncertainty quantification while maintaining scalability.

Moving forward, there are several potential extensions and avenues for future research.
One promising direction is the incorporation of a temporal component, enhancing the proposed methodology's utility in climate model emulation and capturing temporal variations.
Another useful direction to enhance the flexibility and applicability of the MVTM approach is the inclusion of covariates.
Here, the distribution of $\by$ is related to covariate values. In the context of climate models, this would allow to interpolation between emission scenarios.

A continuation of the extension presented is to make the model highly multivariate.
While providing most flexibility, the current formulation requires relating $\binom{P}{2}$~entries in the hyperparameter to the process positions.
This approach is suitable for relatively small $P$.
However, it seems prohibitive when considering all 1168 variables in the CESM.

Furthermore, an important extension to consider is relaxing the assumption of conditional normality at each location.
By allowing for more flexible modeling, such as accounting for skewed or heavy-tailed data, we can extend the Bayesian transport map to multivariate spatial fields where one field may represent extreme values.

In summary, the MVTM approach offers a powerful tool for learning the distribution of multivariate spatial fields. Its ability to handle non-linearity, scalability, and potential for future enhancements make it a promising methodology in spatial statistics.

\footnotesize
\appendix

\subsubsection*{Conflict of interest}
The authors report no conflict of interests.

\subsubsection*{Acknowledgments}

PW and MK were supported by NASA's Advanced Information Systems Technology Program (AIST--21). MK was also partially supported by National Science Foundation (NSF) Grant DMS--1953005. 
\\[5pt]
We would like to thank Jonathan Hobbs for helpful comments and discussions as well as Daniel Drennan for spotting typos in the manuscript and suggestion upon its improvement.

\bibliographystyle{apalike}
\bibliography{mendeley,additionalrefs}

\clearpage %
\section{Algorithms \label{app:algorithm}}

\vspace{-1.5\baselineskip}
\begin{algorithm}[h]
\footnotesize
\caption{Pseudo code of the algorithm estimating the hyperparameters. The training data $\bY$ and validation data $\bV$ contains the ordering and conditioning sets. When the estimation mode is CPP, the elements of $\bftheta$ associated with the latent positions are not updated. The epochs after which a reordering happens in estimation mode OR are predefined.}\label{alg:cap}

\KwData{Training data $\bY$, validation data $\bV$, initial hyper parameters $\bftheta$, batchsize $bs$, learning rate $lr$, max epochs $E$, patience, estimation mode}
\KwResult{Estimated hyperparameter $\hat{\bftheta}$}
\BlankLine
best\_logilik $\gets -\infty$\\
$\hat{\bftheta} \gets \bftheta$\\
patience\_counter $\gets$ 0\\
\tcc{epoch loop}
\For{epoch $\leftarrow$ 0 \KwTo $E$}{
	batches $\gets$ create batches of $\bY$ of size $bs$\\
	\tcc{for each batch, perform gradient based update}
	\For{batch in batches}{
		$\bftheta = \bftheta + lr \nabla_{\bftheta} \log(\operatorname{int\_lik}(batch, \bftheta))$\\
	}
	\tcc{update patience counter; break if patience is exhausted}
	\eIf{$\log(\operatorname{int\_lik}(\bV, \bftheta)) > \text{best\_logilik}$}{
		best\_logilik $\gets \log(\operatorname{int\_lik}(\bV, \bftheta))$\\
		$\hat{\bftheta} \gets \bftheta$\\
		patience\_counter $\gets$ 0\\	
	}{
		patience\_counter $\gets$ patience\_counter + 1\\
	}
	\If{\text{patience\_counter} $>$ \text{patience}}{
		break loop
	}
	\tcc{check if reordering is required and perform if so}
	\If{mode is OR and epoch indicates reordering}{
		$\bY \gets$ update ordering and conditioning sets of $\bY$ using current positions $\bftheta$\\
		$\bV \gets$ update ordering and conditioning sets of $\bV$ using current positions $\bftheta$\\
		\tcc{reset patience counter}
		best\_logilik $\gets \log(\operatorname{int\_lik}(\bV, \bftheta))$\\
		$\hat{\bftheta} \gets \bftheta$\\
		patience\_counter $\gets$ 0\\
	}	
}
\Return $\hat{\bftheta}$
\BlankLine
\end{algorithm}

\end{document}